\documentstyle[epsfig,aps,floats,preprint]{revtex}

\newcommand{\beq}{\begin{equation}}
\newcommand{\eeq}{\end{equation}}
\newcommand{\bea}{\begin{eqnarray}}
\newcommand{\eea}{\end{eqnarray}}

\def\laq{~\raise 0.4ex\hbox{$<$}\kern -0.8em\lower 0.62
ex\hbox{$\sim$}~}
\def\gaq{~\raise 0.4ex\hbox{$>$}\kern -0.7em\lower 0.62
ex\hbox{$\sim$}~}

\def \la {\lambda}

\def \Da {\Delta}

\def \a {\alpha}

\def \ga {\gamma}

\def \da {\delta}

\def \r {\rho}

\def \Om {\Omega}

\def \ti {\tilde}

\begin{document}
\par
\begingroup

\begin{flushright}
BA-TH/00-374\\
January 2000\\
gr-qc/0001094\\
\end{flushright}
\vskip 1true cm

\vspace{10mm}
{\large\bf\centering\ignorespaces
Signal-to-noise ratio for a  stochastic background\\
of massive relic particles
\vskip2.5pt}

\bigskip
{\dimen0=-\prevdepth \advance\dimen0 by23pt
\nointerlineskip \rm\centering
\vrule height\dimen0 width0pt\relax\ignorespaces
M. Gasperini
\par}

{\small\it\centering\ignorespaces
Dipartimento di Fisica, Universit\`a di Bari, \\
Via G. Amendola 173, 70126 Bari, Italy \\
and \\Istituto Nazionale di Fisica Nucleare, Sezione di Bari,
Bari, Italy \\
\par}

\par
\bgroup
\leftskip=0.10753\textwidth \rightskip\leftskip
\dimen0=-\prevdepth \advance\dimen0 by17.5pt \nointerlineskip
\small\vrule width 0pt height\dimen0 \relax

\begin{abstract}
We estimate the signal-to-noise ratio for two gravitational detectors
interacting  with a stochastic background of massive scalar waves.
We find that the present experimental level of sensitivity could be
already enough to detect a signal from a light but non-relativistic
component of dark matter, even if the coupling is weak enough to
exclude observable deviations from standard gravitational
interactions, provided the mass is not too far from the sensitivity and
overlapping band of the two detectors. 
\end{abstract}

\vspace{5mm}

\par\egroup
\thispagestyle{plain}
\endgroup



The sensitivity of present detectors to a stochastic background of relic
gravitational waves has been recently discussed in detail in many
papers (see \cite{1} -\cite{4}, for instance, and references therein). The
sensitivity analysis has also been extended to include scalar waves
\cite{5}, and scalar stochastic backgrounds \cite{6} of massless (or
massive, but light enough) scalar particles, interacting with
gravitational strength with the detectors. At present, however, no
analysis seems to be available on the possible response of the
gravitational antennas to a scalar stochastic background of {\em
non-relativistic} particles. 

The aim of this paper is to compute the signal-to-noise ratio  (SNR) for a
pair of gravitational antennas by taking into account the possible mass
of the  background particles, in order to discuss in some detail the
possible effects of the non-relativistic branch of their spectrum. 

We shall consider a cosmic stochastic background of massive scalar
waves, whose energy density is coupled to the total mass of the
detector with gravitational strength (or weaker). We shall assume that
the background is characterized by a spectral energy density $
\Om(p)= d (\r/\r_c)/d\ln p$, which we measure in  units of critical
density $\r_c= 3H_0^2M_p^2/8\pi$, and which extends in momentum
space from $p=0$ to $p=p_1$ ($p_1$ is a cut-off scale depending on the
details of the production mechanism). As a function of the frequency
$f=E(p)=(m^2+p^2)^{1/2}$, the spectrum $\ti \Om (f)$, 
\beq
\ti \Om (f) \equiv {d (\r/\r_c)\over d\ln f} =
\left(f\over p\right)^2 \Om(p) 
\label{1}
\eeq
thus extends over frequencies $f \geq m$, from $f=m$ to $f=f_1= 
(m^2+p_1^2)^{1/2}$ (note that we are using ``unconventional" units in
which $h=1$, for a better comparison with the observable quantities
used in the experimental analysis of gravitational antennas). We may
thus distinguish three phenomenological possibilities. 

\begin{itemize}

\item{} $m \gg f_0$,  where  $f_0$ is any frequency in the sensitivity
band of the detector (tipically, if we are considering resonant masses
and interferometers, $f_0 \sim 10^2-10^3$ Hz). In this case we expect
no signal, as the response to the background should be totally
suppressed by the intrinsic noise of the detector.

\item{} $m \ll f_0$. In this case the detector, in its sensitivity band,
responds to a relativistic frequency spectrum, and the SNR can be easily
estimated by using the standard  results.  For a
relativistic background of cosmological origin, however, the maximal
amplitude allowed by nucleosynthesis \cite{7} is $\Om \sim 10^{-5}$,
possibly suppressed by a factor $q^2 \ll 1$ (in the interaction with the
antenna) to avoid scalar-induced, long-range violations of the
equivalence principle (see  \cite{8}, for instance). We thus expect from
such a scalar background a response not larger than from a background
of relic gravitons, and then too weak for the sensitivity of present
detectors. 

\item{} $m \sim f_0$. In this case the mass is the frequency band of
maximal sensitivity,  and the detector can respond {\em resonantly} 
also to the {\em non-relativistic part} of the background (i.e. to the
branch $p<m$ of $\Om(p)$).  In the non-relativistic sector, on the other
hand, the background amplitude is not constrained by the
nucleosynthesis bound, because the non-relativistic energy density
grows in time with respect to the relativistic one: it could be
sub-dominant at the nucleosynthesis epoch, even if today has reached
a near-to-critical amplitude $\Om \sim 1$ (i.e., even if the massive
background we are considering represents today a significant fraction
of the cosmological dark matter). In such case, it will be shown in this
paper that the present sensitivity of the existing gravitational
antennas could be  enough to distinguish the physical signal
from the intrinsic experimental noise. 
\end{itemize}

We will follow the standard approach (see \cite{2}, for instance) in
which the outputs of two detectors, $s_i(t)$, $i=1,2$, are correlated
over an integration time $T$, to define a signal: 
\beq
S = \int_{-T/2}^{T/2} dt~ dt' s_1(t)s_2(t') Q(t-t'). 
\label{2}
\eeq
Here $Q(t)$ is a real ``filter" function, determined so as to optimize the
signal-to-noise ratio (SNR), defined by an ensemble average as:
\beq
SNR= \langle S \rangle /\Da S \equiv \langle S \rangle
\left(\langle S^2 \rangle- \langle S \rangle^2\right)^{-1/2}
\label {3}
\eeq
The outputs $s_i(t)= h_i(t)+n_i(t)$ contain the physical strain induced by
the cosmic background, $h_i$, and the intrinsic instrumental noise,
$n_i$. The two noises are supposed to be uncorrelated (i.e., statistically
independent), $\langle n_1(t)n_2(t')\rangle=0$, and much larger  in
magnitude than the physical strains $h_i$. Also, the cosmic background
is assumed to be isotropic, stationary and Gaussian, with $\langle
h_i\rangle=0$. It follows that:
\beq
\langle S\rangle = \int_{-T/2}^{T/2} dt ~dt'\langle h_1(t)h_2(t')\rangle
Q(t-t').  
\label{4}
\eeq

An explicit compuation of the strain, at this point, would require a
specific model of the interaction between the scalar background and
the detector. We will assume in this paper that the strain $h_i(t)$, like
in the case of gravitational waves \cite{2} and Brans-Dicke scalars
\cite{6}, varies in time like the scalar fluctuation  $\phi(x_i,t)$
perturbing the detector (computed at the detector position $x=x_i$),
and is proportional to the so-called ``pattern function" $F_i(\hat
n)=e_{ab}(\hat n)D_i^{ab}$, where $\hat n$ is a unit vector specifying a
direction on the two sphere, $e_{ab}(\hat n)$ is the polarization tensor
of the scalar along $\hat n$, and $D_i^{ab}$ is the detector tensor,
specifying the orientation of the arms of the i-th detector. 

The field $\phi(x,t)$ may represent the scalar (i.e, zero helicity)
component of the metric fluctuations generated by the scalar
component of the background, as in \cite{6}, or could even represent
the background field itself, directly coupled to the detector through a
``scalar charge" $q_i$ (for instance, a dilatonic charge), as discussed in
\cite{9}. To take into account this second possibility, we shall explicitly
introduce the scalar charge in the strain, by setting
\beq
h_i(t)=q_i \phi (x_i,t) e_{ab}(\hat n)D_i^{ab}, 
\label{5}
\eeq
where $q_i=1$ for scalar metric fluctuations, and $q_i<1$ for
long-range scalar fields, phenomenologically constrained by the
gravitational tests. The dimensionless parameter $q_i$ represents the
net scalar  charge per unit of gravitational mass of the detector, and is
in general composition-dependent \cite{9}. 

To compute the average signal (\ref{4}) we now expand the strain in
momentum space,
\bea
&&
h_i(t)= q_i \int dp \int d^2 \hat n ~\phi (p, \hat n) F_i(\hat n)  e^{2\pi
i\left[ p \hat n \cdot {\vec x}_i - E(p) t \right]}, \nonumber \\
&&
p=|\vec p|, ~~~~~~~ \vec p/p = \hat n, ~~~~~~~
E(p)=f=(m^2+p^2)^{1/2}, 
\label{6}
\eea
($d^2\hat n$ denotes the angular integral over the unit two-sphere),
and we use the stochastic condition
\beq
\langle \phi^\star (p, \hat n), \phi(p', \hat n') \rangle= \da (p-p') 
\da^2 (\hat n-\hat n') \Phi (p).
\label{7}
\eeq
The isotropic function $\Phi(p)$ can be expressed in terms of the
spectral energy density $\Om(p)$, defined by
\beq
\r= \r_c \int d\ln p ~\Om (p) = {M_P^2\over 16 \pi }\langle |\dot
\phi|^2\rangle, 
\label{8}
\eeq
($M_P$ is the Planck mass) from which:
\beq
\Phi (p)= {3 H_0^2 \Om (p) \over 8 \pi^3 p E^2(p)}.
\label{9}
\eeq
By inserting the momentum expansion into eq. (\ref{4}), and assuming,
as usual, that the observation time $T$ is much larger than the typical
time intervals $t-t'$ for which $Q\not= 0$, we finally obtain:
\beq
\langle S\rangle= q_1q_2 T {2 H_0^2\over 5 \pi^2} \int {dp\over p
E^2(p)} \ga (p) Q(p) \Om(p). 
\label{10}
\eeq
We have defined the overlap function $\ga(p)$ and the filter function
$Q(p)$, in momentum space, as follows: 
\bea
&&
\ga(p)= {15\over 16 \pi}\int d^2 \hat n F_1(\hat n) F_2 (\hat n) 
 e^{2\pi i p \hat n \cdot ({\vec x}_2 - {\vec x}_1)}, 
\nonumber\\
&&
Q(p)= \int dt' Q(t-t')  e^{2\pi i E(p) (t-t')}. 
\label{11}
\eea
Note that the overlap function depends on the relative distance of the
two gravitational antennas and on their particular geometric
configuration. In the above equation, in particular, $\ga(p)$ has been
normalized  to the response of an interferometric detector to a scalar
wave \cite{6}. 

We need now to compute the variance $\Da S^2$ which, for
uncorrelated noises, much larger than the physical strains,  can be
expressed as \cite{2}:
\beq
\Da S^2 \simeq \langle S^2\rangle = \int _{-T/2}^{T/2} dt dt' d\tau
d\tau' \langle n_1(t) n_1(\tau)\rangle \langle n_2(t') n_2(\tau')\rangle
Q(t-t') Q(\tau-\tau'). 
\label{12}
\eeq
It is convenient, in this context, to introduce the noise power spectrum
in momentum space, $S_i(p)$, defined by 
\beq
\langle n_i(t) n_i(\tau)\rangle = {1\over 2} \int dp S_i(p) 
e^{-2\pi i E(p) (t-\tau)}. 
 \label{13}
\eeq
Assuming, as before, that $T$ is much larger than the typical
correlation intervals $t-t'$, $\tau -\tau'$, and using eq. (\ref{11}) for
$Q(p)$, then yields
\beq
\Da S^2 = {T\over 4} \int {dp\over p} E(p) S_1(p) S_2(p) Q^2(p).
\label{14}
\eeq
The optimal filtering is now determined by the choice (see \cite{2} for
details)
\beq
Q(p)= \la {\ga(p) \Om(p)\over E^3(p) S_1(p) S_2 (p)},
\label{15}
\eeq
where $\la$ is an arbitrary normalization constant. With such a choice
we finally arrive, from eq. (\ref{10}) and (\ref{15}), to the optimized
signal-to-noise ratio:
\beq
SNR = {\langle S\rangle \over \Da S}=   q_1q_2  {4 H_0^2\over 5 \pi^2}
\left[T\int {dp\over p E^5(p)}{\ga^2(p) \Om^2(p)\over  S_1(p)
S_2(p)} \right]^{1/2}. 
\label{16}
\eeq

It must be noted, at this point, that the functions $S_i(p)$ and $\ga(p)$
appearing in the above equation are different, for a massive
background, from the usual noise power spectrum  $\ti S_i(f)$, and
overlap function $\ti \ga(f)$, conventionally used in the experimental
analysis of gravitational antennas. Indeed, $\ti S, \ti\ga$ are defined
as Fourier transforms of the frequency $f=E(p)$, so that (see for
instance eq. (\ref{13})):
\bea
&&
\int df ~\ti S_i(f) e^{-2\pi ift} = \int dp ~ S_i(p) e^{-2\pi iE(p)t}, 
\nonumber\\
&&
\int df ~\ti \ga(f) e^{-2\pi ift} = \int dp ~\ga (p) e^{-2\pi iE(p)t}, 
\label{17}
\eea
from which 
\beq
S_i(p)= (df/dp) \ti S_i(f), ~~~~~~~~~
\ga (p)= (df/dp) \ti \ga(f).
\label{18}
\eeq
By introducing into eq. (\ref{16}) the known, experimentally meaningful
variables $\ti S_i, \ti \ga$, and using  $f= E(p)= (m^2+p^2)^{1/2}$, we
thus arrive at the final expression:
\beq
SNR =  q_1q_2  {4 H_0^2\over 5 \pi^2}
\left[T\int_0^{p_1} {d\ln p\over (m^2+p^2)^{5/2}}{\Om^2(p)~
\ti \ga^2 (\sqrt {m^2+p^2})
\over  \ti S_1(\sqrt {m^2+p^2})\ti  S_2(\sqrt {m^2+p^2})} \right]^{1/2}. 
\label{19}
\eeq

This equation represents the main result of this paper. For any given
massive spectrum  $\Om(p)$, and any pair of detectors with noise $\ti
S_i$ and overlap $\ti \ga$, the above equation determines the range of
masses possibly compatible with a detectable signal ($SNR \gaq 1$), as
a function of their coupling $q_i$ to the detectors. 

For $m=0$ we have $p=f$, and we recover the standard relativistic
result \cite{2}, modulo a different normalization of the overlap
function. For $m \not=0$ we shall assume, as discussed at the
beginning of this paper, that the mass lies within the sensitivity and
overlapping band of the two detectors, i.e. $\ti\ga(m)\not=0$, and
$\ti  S_i(m)$ is near the experimental minimum. Also, let us assume that
the non-relativistic branch of the spectrum, $0<p<m$, is near to
saturate the critical density bound  $\Om <1$, and thus dominates the
total energy density of the background (the contribution of the
relativistic branch $p>m$, if present, is assumed to be negligible). 

To estimate the integral of eq. (\ref{19}), in such case, we can thus
integrate over the non-relativistic modes only. In that range, we will
approximate $\ti S_i$ and $\ti\ga$ with their constant values at $f=m$.
Assuming that the spectrum $\Om(p)$ avoids infrared divergences at $p
\rightarrow 0$ (like, for instance, a blue-tilted spectrum $\Om(p) \sim
(p/p_1)^\da$, with $\da >0$), we define
\beq
\int_0^m d\ln p~ \Om^2(p) = \Om_x^2,
\label{20}
\eeq
where $\Om_x \leq 1$ is a constant,  possibly not very far from unity,
and we finally arrive at the estimate
\beq
SNR \simeq  q_1q_2  {4 H_0^2 \Om_x\over 5 \pi^2}
\left[{T~\ti \ga^2(m)\over  m^5\ti S_1(m)\ti  S_2(m)}
\right]^{1/2}.  
\label{21}
\eeq

Following \cite{2}, the background can be detected, with a detection
rate $\ga$, and a false alarm rate $\a$, if 
\beq
SNR \geq \sqrt 2 \left({\rm erfc}^{-1} 2\a - {\rm erfc}^{-1} 2\ga\right).
\label{22}
\eeq
For a first qualitative indication, let us consider the ideal case in which
the two detectors are coincident and coaligned, i.e. $\ti \ga =1$, $\ti
S_1 =\ti S_2=\ti S$, $ q_1=q_2=q$, and the massive stochastic
background represents a dominant component of dark matter, i.e.
$\Om_x h^2_{100}\sim 1$ (where $h_{100}= H_0/(100~{\rm  km
~sec^{-1}~ Mpc^{-1}})$ reflects  the usual uncertainty in the present
value of the Hubble parameter $H_0$). In such a case eq. (\ref{21}), for
an observation time $T=10^8$ sec, a detection rate $\ga = 95 \%$, a
false alarm rate $\a = 10 \%$, gives the condition:
\beq
m^{5/2} \ti S (m) \laq {q^2\over 3 \pi^2} 10^{-31} {\rm Hz}^{3/2}. 
\label{23}
\eeq

We will use here, for a particular explicit example, the analytical fit of
the noise power spectrum of VIRGO, which in the range from $1$ Hz to
$10$ kHz can be parametrized as \cite{10}:
\bea
\qquad
\ti S(f)&=& 10^{-44} {\rm sec} \Bigg[3.46 \times 10^{-6} \left(f\over
500 ~{\rm Hz}\right)^{-5}+ 6.60 \times 10^{-2}\left(f\over 500~{\rm
Hz}\right)^{-1} \nonumber\\
&+&
3.24\times 10^{-2} + 3.24\times 10^{-2}
\left(f\over 500~{\rm Hz}\right)^{2}\Bigg].
\label{24}
\eea
The intersection of this spectrum with the condition (\ref{23}), in the
plane $\left\{\log \ti S, \log m\right\}$, is shown in Fig. 1 for three
possible values of $q^2$. 
The allowed mass window compatible with  detection  is strongly
dependent on $q^2$, and closes completely for $q^2 < 10^{-7}$, at least
at the level of the noise spectrum used for this example. We should
then consider two possibilities. 

If the spectrum $\Om(p)$ of eq. (\ref{19}) refers to the spectrum of
scalar metric fluctuations, induced on very small sub-horizon scales by
an inhomogeneous, stochastic background of dark matter, then
$q^2=1$ (since the detectors are geodesically coupled to metric
fluctuations). In that case the detectable mass window extends over
the full band from $1$ Hz to $10$ kHz, i.e from $10^{-15}$ to
$10^{-11}$ eV. 

\begin{figure}[t]
\begin{center}
\mbox{\epsfig{file=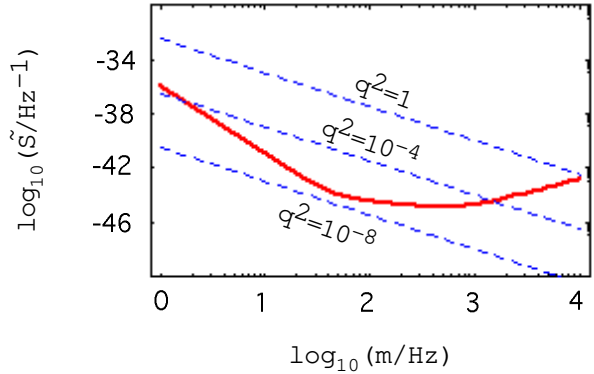,width=82mm}}
\vskip 5mm
\caption{\sl The bold curve corresponds to the noise power spectrum of
VIRGO given in eq. (\ref{24}). The thin, dashed lines represent the
minimal sensitivity required for the detection of the background at the
$90 \%$ confidence level (i.e., $SNR \simeq 2.5$), according to eq.
(\ref{23}). The mass window compatible with detection corresponds to
the range of frequency for which $\ti S$ is below a given dashed line.}
\end{center}
\end{figure}

If, on the contrary, scalar metric fluctuations are negligible on such
small scales, and $\Om(p)$ refers to the spectrum of the scalar
background field itself, directly coupled to the detector through the
scalar charge $q$, then this coupling is strongly suppressed in the
mass range of Fig. 1, which corresponds to scalar interactions in the
range of distance from $10^6$ to $10^{10}$ cm. Otherwise, such scalar
field would induce long range corrections to the standard gravitational
forces that would be detected in the precise tests of Newtonian
gravity and of the equivalence principle (see \cite{11} for a complete
compilation of the bounds on the coupling, as a function of the range). 

Taking into account all possible bounds \cite{11},  it follows that, 
if the scalar coupling is universal (i.e. the induced scalar force is
composition-independent), then  the maximal allowed charge  $q^2$ is
around $10^{-7}$ from $1$ to $10$ Hz, and this upper bound grows
proportionally to the mass (on a logarithmic scale) from $10$ to
$10^{4}$ Hz. Composition-dependent  couplings are instead more
strongly constrained by Eotvos-like experiments: the maximal allowed
value of $q^2$ scales  like in the previous case, approximately, but the
bounds are one order of magnitude stronger. 

By inserting into the condition (\ref{23}) the gravitational bounds on
$q^2$ we are led to the situation illustrated in Fig. 2. A scalar
background of nearly critical density, non-universally coupled to 
macroscopic matter, turns out to be only marginally compatible with
detection (at least, in the example illustrated in this paper), since the
line of maximal $q^2$ is just on the wedge of  the noise spectrum
(\ref{24}). If the coupling is instead universal (for instance, like in the
dilaton model discussed in \cite{8}), but the scalar is not exactly
massless, then there is a mass window open to detection, from
$10^{-14}$ to $10^{-12}$ eV. 

\begin{figure}[t]
\begin{center}
\mbox{\epsfig{file=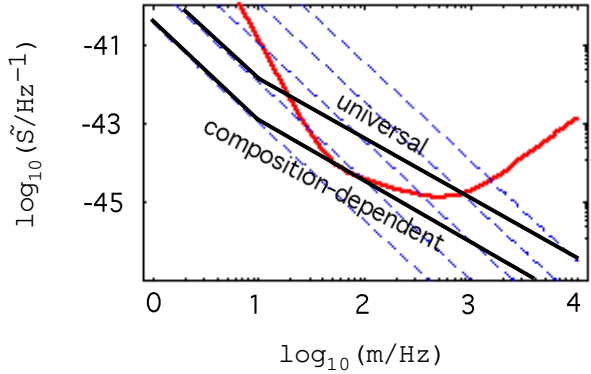,width=82mm}}
\vskip 5mm
\caption{\sl The noise spectrum of Fig. 1 is compared with the maximal
values of $q^2$ (as a function of mass) allowed by gravitational tests,
in two cases: composition-dependent and composition-independent
scalar interactions. The thin dashed lines corresponds, from left to right,
to $q^2=10^{-8}$, $10^{-7}$, $10^{-6}$, $10^{-5}$, $10^{-4}$.  The region
compatible with a detectable signal is above the noise spectrum and
below the bounds given by the gravitational experiments.}
\end{center}
\end{figure}

It seems appropriate to recall, at this point, that it is not impossible to
produce a cosmic background of light, non-relativisic particles that
saturates today the critical energy bound, as shown by explicit
examples of spectra obtained in a string cosmology context \cite{12}.
Such particles, typical of string cosmology, are in general very weakly
coupled to the total mass of the detector (like the dilatons, if they are
long range, and the charge of the antenna is composition-dependent),
or even completely decoupled (like the axions, since the total axionic
charge is zero for a macroscopic, unpolarized antenna).  Nevertheless,
it is important to stress that they could generate a spectrum of scalar
metric fluctuations, gravitationally coupled to the detector, which
follows the same non-relativistic behaviour of the original spectrum.
We know, for instance, that in cosmological models based on the
low-energy string effective action, the variable representing the
dilaton fluctuations exactly coincides with the scalar part of the metric
fluctuations (at least in an appropriate gauge \cite{13}), and that the
associated spectra also coincide. 

In view of the above discussion, the results illustrated in Fig. 1 and Fig.
2 suggest a new possible application of gravitational antennas, which
seems to be interesting. Already at the present level of sensitivity, the
gravitational detectors could be able to explore the possible presence
of a light, massive component of dark matter, in a mass range that 
corresponds to their sensitivity  band, in spite of the fact that such a
massive background could be directly coupled to the total mass of the
detector with a charge much weaker than gravitational, or only
indirectly coupled, through the induced spectrum of scalar metric
fluctuations. 

\acknowledgments
It is a pleasure to thank Michele Maggiore and 
Gabriele Veneziano for interesting and useful discussions.


\begin{references}
\newcommand{\bb}{\bibitem}

\bb{1}E. Flanagan, Phys. Rev. D {\bf 48} (1993) 2389.

\bb{2}B. Allen and J. D. Romano, Phys. Rev. D {\bf 59} (1999) 102001.

\bb{3}M. Maggiore, Phys. Rep. (in press) (gr-qc/9909001).

\bb{4}D. Babusci and M. Giovannini, Phys. Rev. D {\bf 60} (1999) 083511; 
D. Babusci and M. Giovannini, gr-qc/9912035; astro-ph/9912377.

\bb{5}M. Bianchi, M. Brunetti, E. Coccia, F. Fucito and J. A. Lobo,
Phys. Rev. D {\bf 57} (1998) 4525; 
M. Brunetti,  E. Coccia,  V. Fafone and F. Fucito,,  Phys. Rev. D {\bf
59} (1999) 044027.

\bb{6}M. Maggiore and A. Nicolis, gr-qc/9907055. 

\bb{7}V. F. Schwartzaman, JETP Lett. {\bf 9} (1969) 184. 

\bb{8}T. Damour and A. M. Polyakov, Nucl. Phys. B {\bf 423} (1994) 532. 

\bb{9}M. Gasperini, Phys. Lett. B {\bf 470} (1999) 67. 

\bb{10}E. Cuoco, G. Curci and E. Beccaria, in Proc. of the Second E. 
Amaldi Conference on Gravitational Waves (CERN, July 1997), ed. by E.
Coccia et al.,  (World Scientific, Singapore, 1998), p. 524. 

\bb{11}E. Fischbach and C. Talmadge, Nature {\bf 356}
(1992) 207. 

\bb{12}M. Gasperini and G. Veneziano,  Phys. Rev. D  {\bf 59}  (1999)
43503.

\bb{13}R. Brustein, M. Gasperini, M. Giovannini, V. F. Mukhanov and G.
Veneziano,  Phys. Rev. D  {\bf 51}  (1995) 6744.

\end{references}
\end{document}